\documentclass[%
 reprint,
showpacs,
 amsmath,amssymb,
 aps,
 prl,
longbibliography,
floatfix
]{revtex4-1}

\usepackage[utf8]{inputenc}	
\usepackage[T1]{fontenc}	
\usepackage{graphicx}		
\usepackage{color}
\usepackage[english]{babel}
\usepackage{hyperref}
\usepackage{url}
\usepackage{todonotes}
\usepackage{lipsum}
\usepackage{bm}
\usepackage{microtype}

\usepackage{braket}

\usepackage[position=top,caption=false]{subfig}

\let\originaleqref\eqref
\renewcommand{\eqref}{Eq.~\originaleqref}

\newcommand{\fref}[1]{\figurename~\ref{#1}}

\newcommand{\e}[1]{\mathrm{e}^{#1}}

\renewcommand{\L}[0]{\hat{L}}

\renewcommand{\b}{\hat{b}}
\renewcommand{\c}{\hat{c}}
\renewcommand{\a}{\hat{a}}

\renewcommand{\H}{\hat{H}}

\graphicspath{
{..figures/}
{/home/alexander/Dropbox/PhD/generalizedInputOutput/figures/}
}

\begin{document}
\title{Input-Output Theory with Quantum Pulses}
\author{Alexander Holm Kiilerich}
\email{kiilerich@phys.au.dk}
\author{Klaus Mølmer}
\email{moelmer@phys.au.dk}
\date{\today}
\affiliation{Department of Physics and Astronomy, Aarhus University, Ny Munkegade 120, DK 8000 Aarhus C. Denmark}
\date{\today}

\bigskip

\begin{abstract}
We present a formalism that accounts for the interaction of a local quantum system such as an atom or a cavity with travelling pulses of quantized radiation. We assume Markovian coupling of the stationary system to the input and output fields and non-dispersive asymptotic propagation of the pulses before and after the interaction. This permits derivation of a master equation where the input and output pulses are treated as single oscillator modes that both couple to the local system in a cascaded manner. As examples of our theory we analyse reflection by an empty cavity with phase noise, stimulated atomic emission by a quantum light pulse, and formation of a Schrödinger-cat state by the dispersive interaction of a coherent pulse and a single atom in a cavity.
\end{abstract}

\maketitle
\noindent

\paragraph{Introduction.---}
Quantum states of light may find applications for precision sensing \cite{schnabel2010quantum,wolfgramm2013entanglement} and as processing or flying qubits in quantum computers and quantum communication networks \cite{kimble2008quantum,o2009photonic}. While the intuition behind generation of single photon and multi-photon states and demonstration of atom-photon and photon-photon quantum interactions portrays the state of light as a superposition of Fock states of a single mode or a few modes, propagating fields in reality explore an infinite continuum of modes which prohibit a full quantum treatment by a Schrödinger picture wave function or density matrix.

If the physical setup contains guided fields and material systems with only a single shared quantum of excitation, the quantum state can be expanded on discrete excited states and single quantum wave packets (see, e.g., \cite{Motzoi_2018}). The introduction of further quanta of energy, however, complicates matters significantly, as both particle aspects (photon number) and wave packet aspects require a full quantum treatment 
(for a recent review of theory approaches see Ref.~\cite{fischer2018scattering}).
While expansion of the field state on the continua of one and two-photon states \cite{PhysRevA.92.033803, PhysRevA.82.063821} may be adequate to describe many processes relevant to quantum information processing \cite{witthaut2010photon,witthaut2012photon,nisbet2013photonic,PhysRevA.82.033804,PhysRevA.82.033804,PhysRevLett.114.173601,bock2018high}, a more general and more tractable theory is desired.
Indeed, Itô calculus approaches \cite{gheri1998photon,gough2012single} and the so-called Fock master equation \cite{PhysRevA.86.013811} permit evaluation of the state of a quantum system which is driven by an incident quantum pulse in a superposition of Fock states. Mean values and correlation functions of the fields emitted by the system can then be expressed in terms of system observables, but they do not provide the full quantum state of the output field.

\begin{figure}[h!]
\centering
\includegraphics[trim=0 0 0 0,width=0.8\columnwidth]{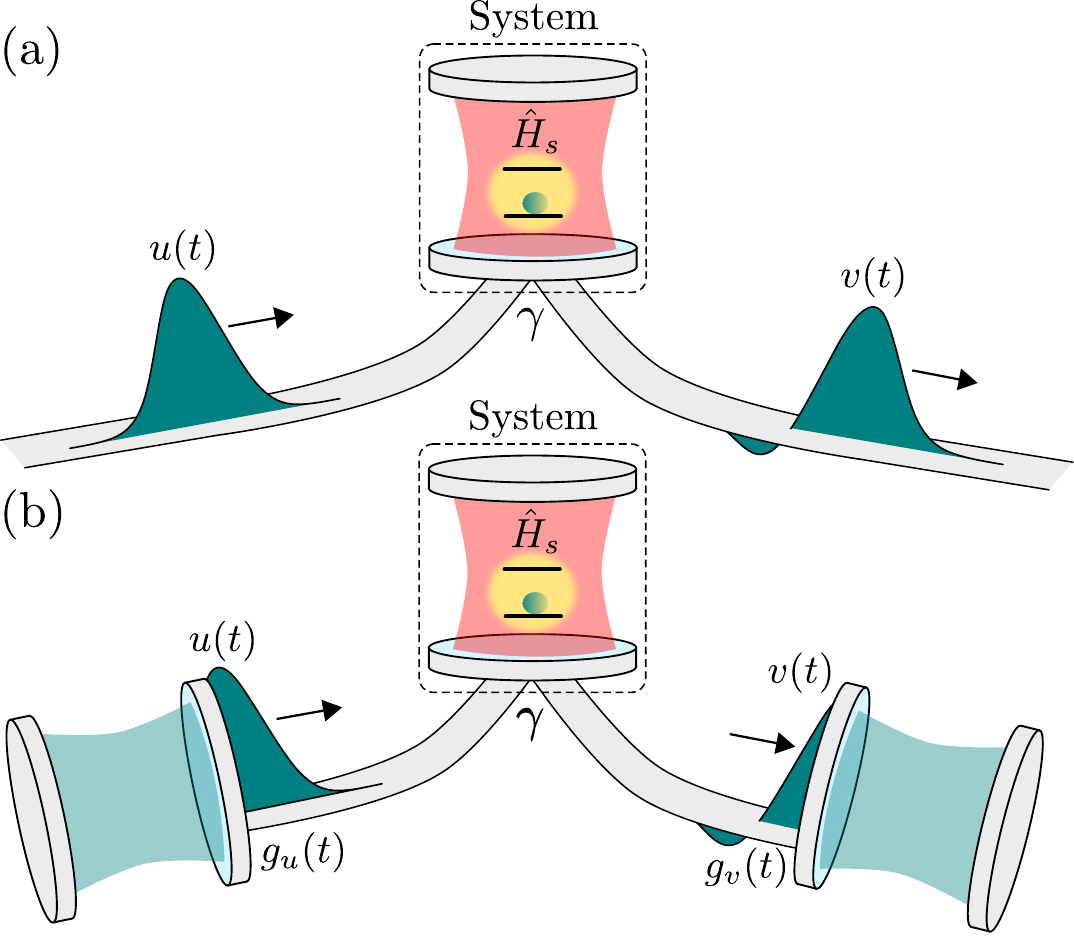}
\caption{(a): An incident pulse with a temporal envelope $u(t)$ and arbitrary quantum state content interacts with strength $\sqrt{\gamma}$ with an arbitrary quantum system observable $\hat{c}$. The system is described by a Hamiltonian $\H_s$, and it is depicted here as a two-level system inside a resonator. In this letter we provide a theory for the full quantum state of light occupying any  reflected temporal pulse $v(t)$.  (b): We model the situation of arbitrary travelling pulses in (a) by virtual cavities with complex, time dependent mirror-couplings, $g_u(t)$ and $g_v(t)$, designed so they exactly eject and absorb the modes $u(t)$ and $v(t)$.}
\label{fig:setup}
\end{figure}

The emission from a quantum system will not in general be restricted to a single mode, but we can choose to examine any particular propagating wave packet and consider the quantum state occupying just that mode after the interaction with the quantum system.
Our theory thus accounts for the kind of experiment depicted in \fref{fig:setup}(a), where a wave packet is incident on an arbitrary quantum system, which we assume can be adequately described in a Hilbert space of finite dimension $d$. The quantum state of a suitably defined outgoing wave packet is precisely the information retained by typical quantum communication or computing protocols, while the radiation which is not captured by that mode represents loss. Our theory is general and applies for any selected output mode function. At the end of the manuscript we shall propose strategies to select the most relevant, e.g., most populated, mode function and generalizations to deal with multiple input and output pulses.  



\paragraph{Theory.}---
In Figure~\ref{fig:setup}(a), a quantum system described by a possibly time-dependent Hamiltonian  $\H_{\mathrm{s}}(t)$ is coupled to an input bosonic field $\b_{\mathrm{in}}(t)$ by an interaction $(\hbar = 1)$ $\hat{V}_{SB}(t) = i \sqrt{\gamma}[(\c \b_{\mathrm{in}}^\dagger(t)-\c^\dagger \b_{\mathrm{in}}(t)]$ where $\c$ is a system operator. If $\c$ is a lowering operator, $\gamma$ is the corresponding decay rate of excitations in the system, and the outgoing field after interaction with the system is given by the input-output operator relation, $\b_{\mathrm{out}}(t) = \b_{\mathrm{in}(t)}+\sqrt{\gamma}\c(t)$ \cite{PhysRevA.31.3761,GardinerBook}. Direct application of this expression requires knowledge of the time dependent system operator $\c(t)$ in the Heisenberg picture which is only available if $\H_{\mathrm{s}}(t)$ is sufficiently simple (e.g., quadratic in oscillator raising and lowering operators $\c$ and $\c^\dagger$ \cite{GardinerBook}). 

Since we shall treat the case of a quantum state input occupying a single normalized wave packet $u(t)$, it is natural to seek a Schr\"odinger picture description of the input by the Fock states $|n\rangle$, related to a single bosonic creation operator
\begin{align}\label{eq:x}
\b_u^\dagger = \int dt\, u(t) \b^\dagger(t).
\end{align}
The pulse shape is modified by the interaction and the outgoing pulse may acquire multi-mode character, which complicates a full numerical treatment. However, it is possible to consider the output radiation from the system, carried by any particular outgoing mode function $v(t)$, as sketched  in \fref{fig:setup}(a). The essential idea of our approach is therefore to describe the input and output pulses by two separate field modes. Assuming the Born-Markov approximation, this can be done in an exact manner.  

To alleviate the problem of dealing with the spatio-temporal propagation of quantum fields, we note that any arbitrary wave packet can be emitted as the output from - or absorbed as the input to - a virtual one-sided cavity with time-dependent complex coupling
$i[g^*(t)\a\b_{\mathrm{in}}^\dagger-g(t)\a^\dagger \b_{\mathrm{in}}]$ to its input continuum fields. These virtual cavities work as coherent beam-splitters between the discrete intra-cavity modes and specific wave packets incident on and emanating from the cavities. In particular, if $g_u(t)$ is chosen such that 
\begin{align}\label{eq:g1}
g_u(t)  = \frac{u^*(t)}{\sqrt{1-\int_0^t dt'\, |u(t')|^2}},
\end{align}
the initial intracavity quantum state at $t=0$ is emitted as a travelling wave packet  given by the time dependent  mode function $u(t)$ \cite{gough2015generating}.  An alternative protocol to release a cavity state into a specific complex wave packet, applying a real coupling coefficient and a time dependent cavity detuning, was derived in \cite{gheri1998photon}.

Similarly, a single mode cavity with complex input coupling
\begin{align}\label{eq:g2}
 g_v(t) = -\frac{v^*(t)}{\sqrt{\int_0^t dt'\, |v(t')|^2}}.
\end{align}
will asymptotically acquire the quantum state content of a wave packet $v(t)$ incident on the cavity. This result is readily shown by the equivalent equations for classical field amplitudes and for single photon wave packets \cite{7798639}, 

Rather than propagating pulses interacting with a local scatterer, we can thus describe the problems as a cascaded system with time dependent couplings, see \fref{fig:setup}(b). Due to the  
assumption of Markovian coupling to the continuous field degrees of freedom and dispersion free propagation of the wave packets, we can apply the cascaded system analysis by Gardiner \cite{PhysRevLett.70.2269} and Carmichael \cite{PhysRevLett.70.2273} to obtain a master equation that involves only the quantum states of the intermediate quantum system and the field states of the two cavity modes, represented by field operators $\a_u$ and $\a_v$.

This can be accomplished in a systematic manner in the so-called SLH framework \cite{gough2009series,doi:10.1080/23746149.2017.1343097}, by concatenating the Hamiltonians and damping terms according to the routing of output from one system into another.
The density matrix of the total system evolves according to a master equation on the Lindblad form,
\begin{align}\label{eq:me}
\frac{d\rho}{dt} = -i[\H,\rho]+ \sum_{i=0}\left( \L_i\rho \L_i^\dagger-\frac{1}{2}\left\{\L_i^\dagger \L_i, \rho\right\}\right),
\end{align}
where $\{\cdot,\cdot\}$ denotes the anticommutator, and the Hamiltonian
\begin{align}\label{eq:H}
\begin{split}
&\H(t) = \H_{\mathrm{s}}(t)+\frac{i}{2}\big(\sqrt{\gamma}g_u^*(t)\a_u^\dagger\c
\\
&+ \sqrt{\gamma^*} g_v(t)\c^\dagger\a_v  + g_u^*(t) g_v(t)\a_u^\dagger\a_v - \mathrm{h.c.}\big)
\end{split}
\end{align}
contains terms that represent coherent exchange of energy between the three components.

The damping terms in \eqref{eq:me} include a single Lindblad operator of the form.
\begin{align}\label{eq:L}
\L_0(t) = \sqrt{\gamma}\c+g_u(t)\a_u+g_v(t)\a_v,
\end{align}
representing the output loss from the last cavity, as well as operators $\L_i$ with $i>0$, representing separate decay and loss mechanisms of the quantum scatterer. 
The formalism may be extended to include several input and output modes, see supplemental materials \cite{supp}.

The solution to~(\ref{eq:me}) yields the density matrix of the joint system and provides a full quantum state description of the output mode and of its potentially entangled state with the scatterer. Our theory thus goes far beyond the study of expectation values and low order correlation functions of the output field operators.
The restriction of the dynamics from the infinite continuum to only two field modes reduces the infinite dimensional Hilbert space to one of dimension  ${\cal N} \leq (N+1)\times d \times (M+1)$, where $N$ and $M$ are the maximum number of excitations in the incoming and outgoing modes. Our full quantum description amounts thus to the evolution of an ${\cal N}\times {\cal N}$ density matrix $\rho$.
Next, we shall present a few examples of our formalism. Numerical solutions to the master equation~(\ref{eq:me}) are obtained using the QuTiP toolbox \cite{JOHANSSON20121760,JOHANSSON20131234}.

\begin{figure}[]
\centering
\includegraphics[trim=0 0 0 0,width=0.9\columnwidth]{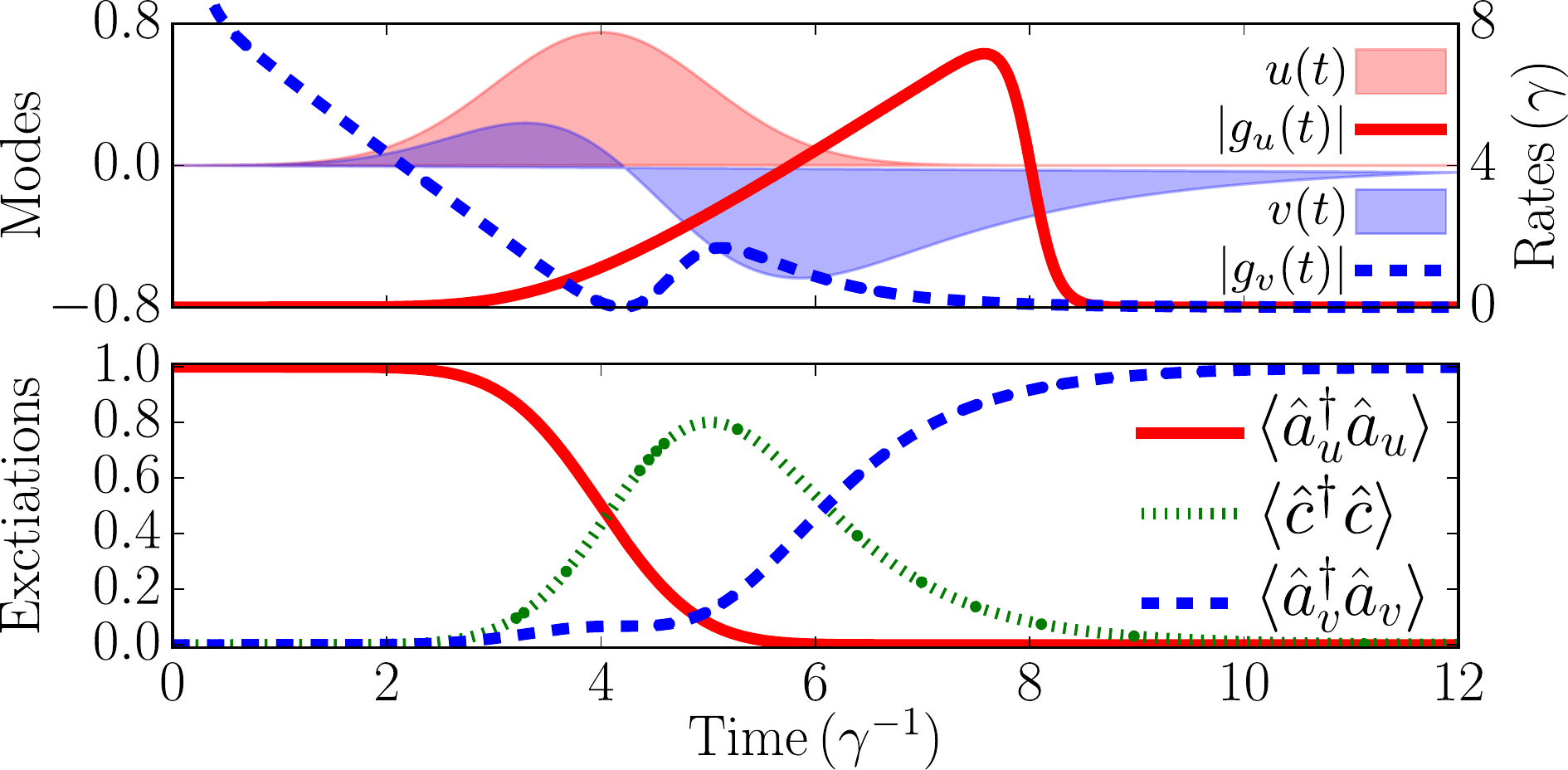}
\caption{
Scattering of a single photon in a Gaussian mode on an empty cavity. Upper panel: The incoming pulse $u(t)$ (red shaded Gaussian shape) and the reflected pulse $v(t)$ (blue shaded non-Gaussian shape), given by (\ref{eq:vGauss}). The  solid (dashed) curves show the squared coupling strengths $|g_u(t)|^2$ and $|g_v(t)|^2$, given by Eqs.~(\ref{eq:g1})~and~(\ref{eq:g2}).
Lower panel: Average photon number as a function of time in the incoming and outgoing  pulses and inside the cavity.
}
\label{fig:emptyCavity}
\end{figure}

\paragraph{Examples.}---
As a first example of our formalism, we consider the scattering on an empty, one-sided cavity with resonance frequency $\omega_c$. The local system Hamiltonian is $\H_s = \omega_c \c^\dagger \c$ and scattering with coupling amplitude $\sqrt{\gamma}$ of the input field to the cavity field is readily described by a frequency dependent reflection coefficient
$
r(\omega) = [i(\omega-\omega_c)+\gamma/2]/[i(\omega-\omega_c)-\gamma/2].
$
That is, the Fourier transformed pulse shapes obey
\begin{align}\label{eq:vGauss}
v(\omega) = r(\omega)u(\omega).
\end{align}
In the upper panel of \fref{fig:emptyCavity}, we show how a real Gaussian pulse $u(t)$ is reflected into a mode $v(t)$ which is also real but has a sign change around the time $\gamma t  = 4$. The squared value of the corresponding coupling strengths $|g_{u(v)}(t)|^2$  are shown in the same panel. The lower panel shows how the average photon number in the input, cavity and output modes change with time for an initial one-photon Fock state in the input pulse. We emphasize that in this case the perfect state transfer is guaranteed to the \textit{known} output mode. If we solve the master equation (\ref{eq:me}) with any other choice of output mode, the transfer will be imperfect.

As an example of a system that scatters a single input pulse into a multi mode output, we consider phase noise in the system, e.g., due to a jittering of one of the cavity mirrors on a timescale $\tau_{\mathrm{jit}}$. This imposes an additional Lindblad term $\L_1 = \tau_{\mathrm{jit}}^{-1/2} \c^\dagger \c$ in the master equation~(\ref{eq:me}) (see Ref.~\cite{PhysRevA.85.013844} for an extended discussion of this model) but poses no problem for our numerical solution of the problem.

\begin{figure}[h!]
\centering
\includegraphics[trim=0 0 0 0,width=0.75\columnwidth]{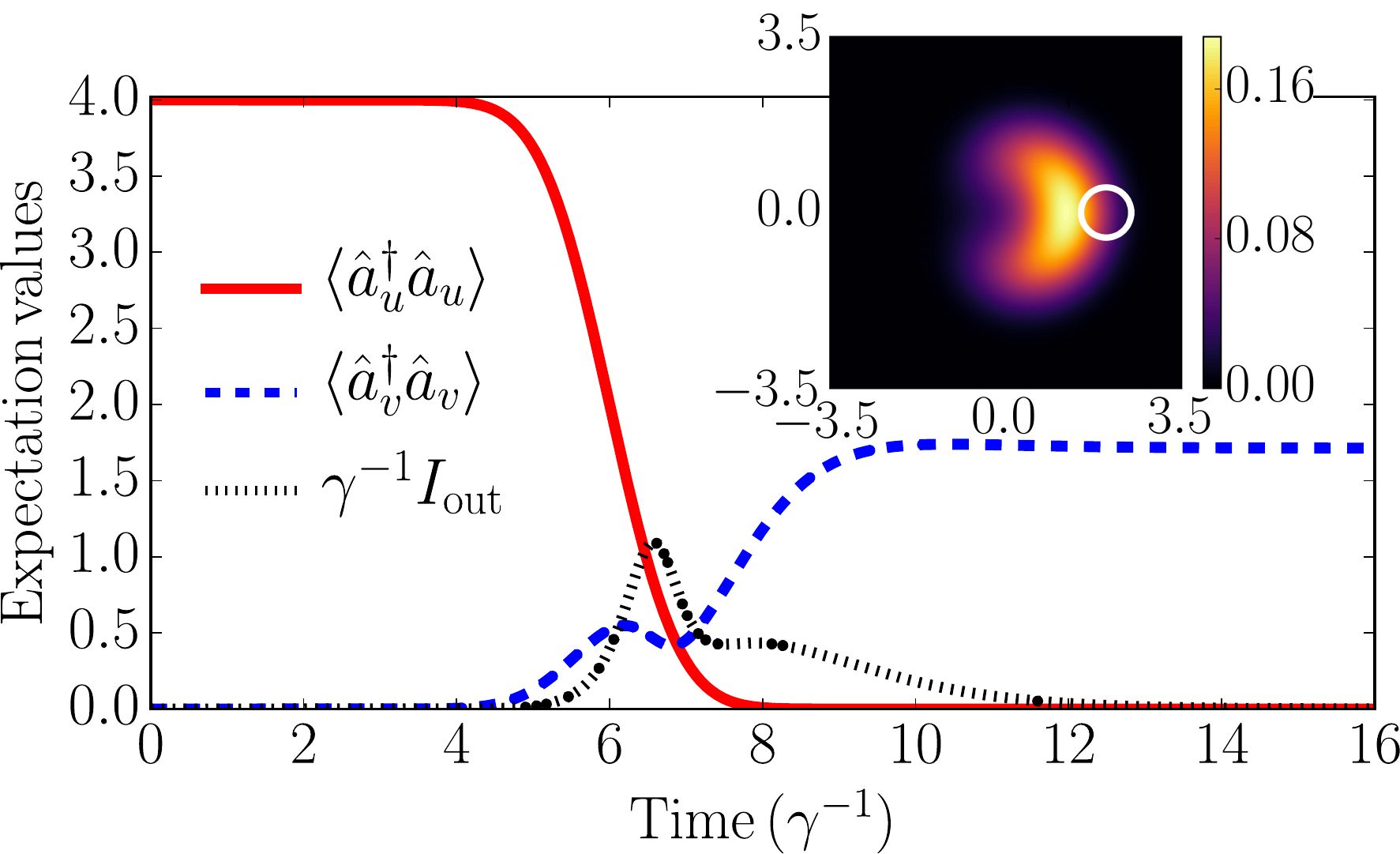}
\caption{Scattering on an empty cavity with phase noise.
Main plot: Average photon number in the ingoing and outgoing modes and the flux $I_\mathrm{out}$ (in units of $\gamma$) from the final virtual cavity as functions of time. Inset: Wigner function of the outgoing mode at the final time $\gamma t = 10$ with field quadratures $\mathrm{Re}(\braket{\a_v})$ and $\mathrm{Im}(\braket{\a_v})$ shown along the x and y axes, respectively.
The circle illustrates the phase space location and width of the initial coherent state in the incoming mode $u(t)$.
Results are provided for the modes $u(t)$ and $v(t)$ shown in the upper panel of \fref{fig:emptyCavity} and for an incident coherent state $|\alpha = 2\rangle$ and $\tau_{\mathrm{jit}} = \gamma^{-1}$.
}
\label{fig:phaseNoise}
\end{figure}

In \fref{fig:phaseNoise}, we present calculations for the same input and output modes as in \fref{fig:emptyCavity} with the  incoming pulse prepared in a coherent state $\ket{\alpha = 2}$. The phase noise causes an imperfect transfer to the examined output mode $v(t)$ and a corresponding output flux $I_{\mathrm{out}}(t) = \langle \hat{L}_0^\dagger \hat{L}_0\rangle$ from the final virtual cavity at intermediate times.
The insert Wigner function shows that the quantum state of the outgoing mode is not a coherent state but may be described as a statistical mixture of coherent states with reduced amplitude and rotated by a range of different complex phases.

Through a comprehensive derivation, relying on Itô calculus, Fischer \cite{Fischer:18} has investigated the emission from an excited atom, stimulated by an incident quantum pulse and particularly how efficiently such stimulated emission occurs into the mode occupied by the incident photons. Our formalism allows treatment of this problem with minimal effort. Imagine a two level atom with ground state $\ket{g}$ which is prepared in its excited state $\ket{e}$ and decays at a rate $\gamma$ by the dipole lowering operator $\c = \ket{g}\bra{e}$.

\begin{figure}[]
\centering
\includegraphics[trim=0 0 0 0,width=0.85\columnwidth]{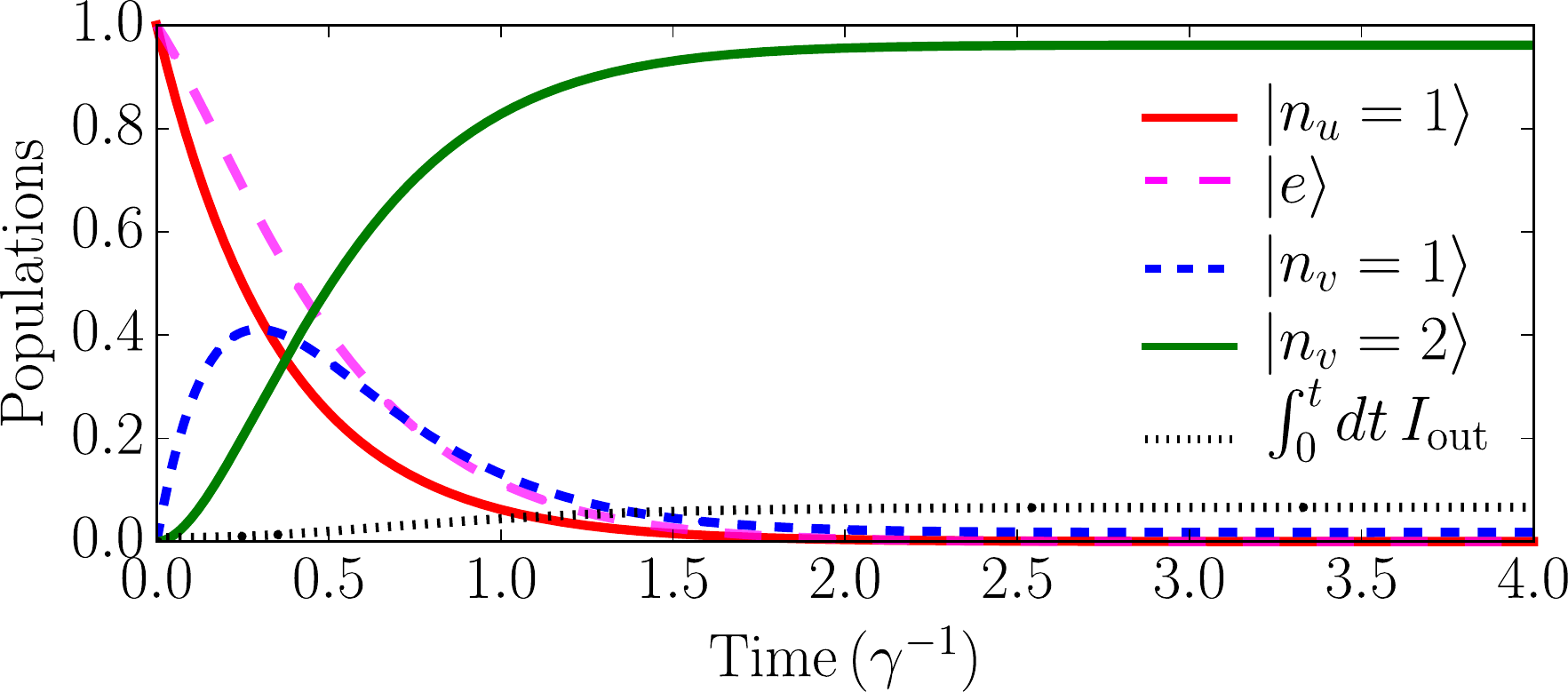}
\caption{Stimulated emission. The figure shows the time evolution of the populations in the one photon components of the incoming pulse, the excited state of the atomic emitter and the one and two photon components of the outgoing pulse. The dotted curve shows the number of excitations $\gamma^{-1}\int_0^t dt \, I_{\mathrm{out}}$ lost into other modes.
}
\label{fig:stimEmi}
\end{figure}

An exponentially decaying mode $u(t) = \sqrt{\Gamma}\e{-\Gamma t/2}\Theta(t)$, where $\Theta(t)$ is the Heavyside step function, has been identified as optimal for stimulated emission \cite{Valente_2012} where the optimal value of $\Gamma$ depends on the quantum state of the incoming pulse. The fiducial cavity couplings leading to this ansatz for $u(t)$ and $v(t)$ are given by Eqs.(\ref{eq:g1})~and~(\ref{eq:g2}) as $g_u(t) = \sqrt{\Gamma }\Theta(t)$ and $g_v(t) = \sqrt{\Gamma/(\e{\Gamma t}-1)}\Theta(t)$.
For an incident one photon Fock state for which the optimal value is $\Gamma = \gamma/0.36$, the interaction with the excited atom causes the outgoing mode $v(t)$ to first acquire a one-photon component which is gradually replaced by a two photon component with a final population of $0.97$, see \fref{fig:stimEmi}. This confirms that stimulated emission has indeed occurred, but due to a minute mode mismatch, around 0.07 photon ($=\int dt \, I_{\mathrm{out}}$) is lost to orthogonal modes.

\begin{figure}[]
\centering
\includegraphics[trim=0 0 0 0,width=1\columnwidth]{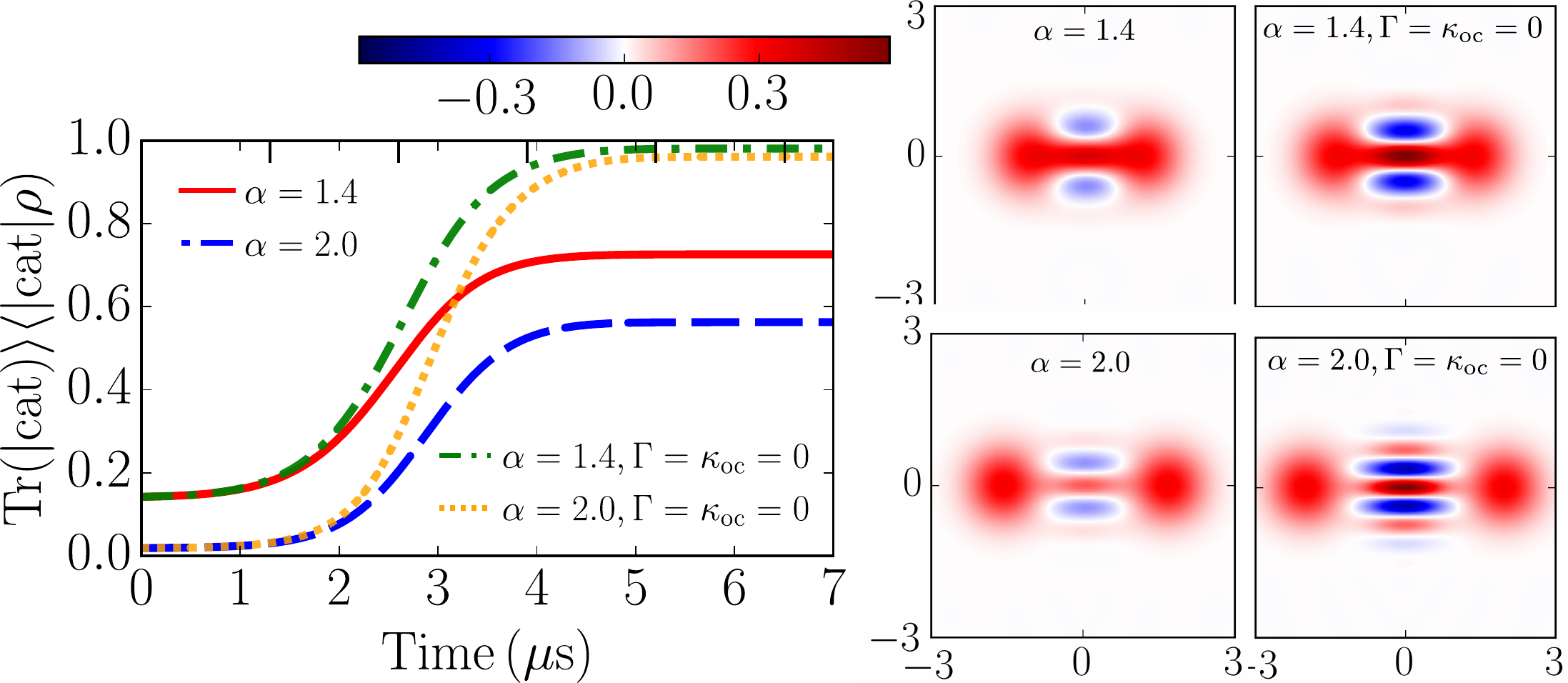}
\caption{
Creation of a Schrödinger cat. Results are shown for $\alpha = 1.4$ and $2.0$ and assume the parameters given in Ref.~\cite{hacker2019deterministic}, $(g,\gamma,\Gamma,\kappa_{\mathrm {oc}}) = 2\pi\times(15.6,4.6,6.0,0.4)$MHz.
Left panel: Fidelity with the state~(\ref{eq:cat}) as a function of time. For comparison, we show also the fidelity, assuming ideal conditions with no decoherence ($\Gamma=\kappa_{\mathrm{oc}} = 0$).
Color plots: Wigner function of the outgoing pulse after a $\pi/2$ spin rotation and post selection on detection of the atom in the state $\ket{\downarrow}$. According to \eqref{eq:cat}, this procedure should ideally produce an even cat state, $(\ket{\alpha}+\ket{-\alpha})/\sqrt{2}$ in the outgoing mode.}
\label{fig:rempeCAT}
\end{figure}

As a final example, we apply our theory to a recent experiment by Hacker $\textit{et al.}$ \cite{hacker2019deterministic} where an atom with two ground states $\ket{\downarrow}$ and $\ket{\uparrow}$ and one excited state $\ket{e}$ is placed inside a cavity with an out-coupling $\gamma$. The transition $\ket{\uparrow}\leftrightarrow \ket{e}$ is strongly coupled to the cavity mode $\c$ by a Hamiltonian $\H_s = g(\ \ket{\uparrow}\bra{e}\c^\dagger+ \ket{e}\bra{\uparrow}\c)$, and an incoming pulse, prepared in a coherent state $\ket{\alpha}$, is reflected with or without a phase shift depending on the state of the atom. If the atom is prepared in a superposition  $(\ket{\downarrow}+\ket{\uparrow})/\sqrt{2}$, one should thus expect the outgoing light pulse to occupy a Schrödinger-cat entangled state with the atom \cite{PhysRevLett.92.127902,RevModPhys.87.1379},
\begin{align}\label{eq:cat}
\ket{\mathrm{cat}} = \frac{1}{\sqrt{2}}(\ket{\uparrow}\ket{\alpha}+\ket{\downarrow}\ket{-\alpha}).
\end{align}
This is verified by our formalism in \fref{fig:rempeCAT} where for $\alpha= 1.4$ we see a 0.98 and for $\alpha = 2$ a 0.96 fidelity with the cat state~(\ref{eq:cat}) at the final time if we assume perfect reflection of a Gaussian mode $v(t) = u(t)$ centered at the time $3\mu$s.
In realistic settings and indeed in Ref.~\cite{hacker2019deterministic}, the fidelity is hampered by atomic decay at a rate $\Gamma$ and leakage of the cavity into other channels at a rate $\kappa_{\mathrm{oc}}$, implying two additional decoherence channels, $\L_1 = \sqrt{\Gamma} \ket{\uparrow}\ket{e}$ and
$\L_2 = \sqrt{\kappa_{\mathrm {oc}}}\c$ in \eqref{eq:me}, as well as imperfections in the matching of the recorded output mode with the actual signal.
The full curve in \fref{fig:rempeCAT} shows how these effects lower the fidelity to 0.72 for $\alpha = 1.4$ while a larger cat state with $\alpha = 2$ suffers more severely, yielding  a fidelity of only $0.56$. The Wigner functions plotted at the final time illustrate, however, that despite the imperfections, the characteristic signature of a Schrödinger cat state emerges in the output pulse, post selected on the atomic state.
Parameters and details concerning the Wigner functions are given in the figure caption.

\paragraph{Finding the optimal output mode(s)}---
Our theory permits evaluation of the quantum state content of any desired output field mode, and as our examples illustrate, an ill-chosen output mode presents a loss and an impediment to retrieve the desired quantum state. We propose to identify the optimal mode function $v(t)$  by first calculating the emitter autocorrelation function $g^{(1)}(t,t')=\langle \L_s^\dagger(t)\L_s(t')\rangle$, where $\L_s(t) = g_u(t)\a_u(t)+\sqrt{\gamma}\c(t)$. This calculation is possible by application of the quantum regression theorem \cite{breuer2002theory,GardinerBook} to the cascaded master equation of the input cavity and quantum system. If the emission occurs into a single mode, the correlation function factorizes and $g^{(1)}(t,t') \propto v^*(t)v(t')$, while in the general case, an expansion $g^{(1)}(t,t') = \sum_i n_i v_i^*(t)v_i(t')$ may be used to identify a few dominant modes with mean photon number $n_i$ for which the output quantum state can be calculated by a few-mode extension of our theory. 

In the supplemental material \cite{supp}, we describe a generalization of our theory which allows a full quantum description of multiple output and input modes. This is achieved by including additional virtual cavities before and after the quantum scatterer in a cascaded fashion.

\paragraph{Outlook.}---
The formalism presented in this letter provides, in a straightforward manner, a full quantum description of a light pulse reflected by a quantum system into one or more distorted modes. Our theory applies equally well to light and other (dispersion free) carriers of quantum states such as microwaves and surface acoustic waves, considered in recent experiments \cite{birnbaum2005photon,schuster2008nonlinear,piro2011heralded,goto2018demand,axline2018demand,
hacker2019deterministic,nisbet2013photonic,bock2018high,satzinger2018quantum} and experimental proposals \cite{PhysRevA.84.033854,PhysRevA.88.033832,PhysRevLett.112.093601,witthaut2012photon,PhysRevLett.118.133601,PhysRevX.7.011035,PhysRevLett.114.173601,PhysRevX.5.031031}.

We illustrated our theory by the solution of the cascaded master equation for the input and output field Fock space density matrices, but the theory may also employ Heisenberg picture and phase space approaches. Similarly, quantum trajectory analyses of heralded or conditional dynamics have been proposed \cite{PhysRevLett.70.2273,PhysRevA.96.023819,fischer2018scattering,Gough_2014} and follow effortlessly from our formalism.

We recall that the time dependent coupling to input and output mode cavities is a purely theoretical construction to arrive at our simple formalism; no such couplings need be implemented in experiments. The chiral coupling and spatial separation of input and output fields in \fref{fig:setup} may be achieved by various means for single sided cavity systems, while two-sided cavities should be described by two (reflected and transmitted) output modes, and more complex interferometric setups with multiple input and output ports may explore an even larger number of modes 
\footnote{The elimination of propagation segments assumes unidirectional (chiral) coupling of the cascaded components, while bidirectional propagation of light with sizeable delays yields non-Markovian effects \cite{lodahl2017chiral}}.


\begin{acknowledgments}
The authors acknowledge support from the European Union FETFLAG program, Grant No. 820391 (SQUARE), and the
U.S. ARL-CDQI program through cooperative Agreement No. W911NF-15-2-0061.
\end{acknowledgments}

%

\end{document}